\documentclass[twocolumn,amsmath,superscriptaddress, longbibliography]{revtex4-1}

\usepackage{graphics}
\usepackage{xcolor}
\usepackage{amssymb,amsfonts,amsmath}
\usepackage{mleftright}
\usepackage{xparse}
\usepackage{amsfonts}
\usepackage{mathtools}
\usepackage{physics}

\def \>{\rangle}
\def \<{\langle}

\newcommand{\bext}{b^{\mathrm{ext}}}
\newcommand{\drive}{\Delta \mu^{\mathrm{drive}}}

\def\be{\begin{equation}}
\def\ee{\end{equation}}
\def\longrightharpoonup{\relbar\joinrel\rightharpoonup}
\def\longleftharpoondown{\leftharpoondown\joinrel\relbar}

\def\longrightleftharpoons{
  \mathop{
    \vcenter{
      \hbox{
      \ooalign{
        \raise1pt\hbox{$\longrightharpoonup\joinrel$}\crcr
	  \lower1pt\hbox{$\longleftharpoondown\joinrel$}
	  }
      }
    }
  }
}

\newcommand \bea {\begin{eqnarray}}
\newcommand \eea {\end{eqnarray}}

\newcommand{\hbAppendixPrefix}{A}

\NewDocumentCommand{\evalat}{sO{\big}mm}{%
  \IfBooleanTF{#1}
   {\mleft. #3 \mright|_{#4}}
   {#3#2|_{#4}}%
}

\begin{document}

\title{Thermodynamic origins of topological protection in nonequilibrium
stochastic systems}

\author{Pankaj Mehta}
\email{pankajm@bu.edu}
\affiliation{Department of Physics, Boston University, Boston, MA 02215, USA}
\affiliation{Biological Design Center, Boston University, Boston, MA 02215, USA}
\affiliation{Faculty of Computing and Data Science, Boston University, Boston, MA 02215, USA}
\author{Jason W. Rocks}
\affiliation{Department of Physics, Boston University, Boston, MA 02215, USA}

\begin{abstract}
Topological protection has emerged as an organizing principle for understanding and engineering robust collective behavior in electronic and material systems. Recent work suggests that topology may also play a role in organizing stochastic processes relevant to biology and self-assembly. Here, we show that topological protection in chemical networks can be understood entirely in terms of nonequilibrium thermodynamics. We illustrate these ideas using simple examples inspired by the literature.
\end{abstract}
\maketitle

Biochemical processes in living systems operate far from equilibrium.
Such nonequilibrium stochastic processes play vital roles in diverse biological settings ranging from information processing to the self-assembly of dynamic structures such as microtubules~\cite{beard2008chemical, qian2005thermodynamics, qian2007phosphorylation, hill2012free, lan2012energy, mehta2012energetic, murugan2012speed, lang2014thermodynamics, mehta2016landauer, ouldridge2017thermodynamics, hess2017non, mcinally2021scaling, das2021chemically}. In all of these phenomena, cells transduce free energy by coupling chemical reactions to high-energy molecules (e.g., ATP and ADP) that are maintained out-of-equilibrium. Outside of biology, similarly driven stochastic chemical processes play a central role in material self-assembly and the design of active materials~\cite{zeravcic2014size, murugan2015undesired, nguyen2016design,bisker2018nonequilibrium}.  In both natural and synthetic systems, it is crucial that the behavior and function of these systems be robust to the presence of disorder and thermal fluctuations.

Recently, topological protection has emerged as an important design principle for understanding and engineering robust behavior in out-of-equilibrium stochastic systems. Inspired by the immense progress made on this topic, several authors have translated many ideas originally developed in the context of quantum mechanics and discrete meta-materials to stochastic networks~\cite{murugan2017topologically, dasbiswas2018topological, knebel2020topological, knebel2020topological, tang2021topology, yoshida2021chiral}. On a technical level, many of these works exploit the formal analogy between the Master Equation and the Schr\"{o}ndiger equation in which the state-to-state transition matrix of the former plays an analogous role to the Hamiltonian of the latter~\cite{van1992stochastic}. As in quantum systems, the existence of a topological phase transition can be related to the closing of a gap in the spectrum (now of the transition matrix instead of the Hamiltonian) and an associated change in the value of a topological invariant, usually a $\mathbb{Z}_2$-valued Zaks phase~\cite{zak1989berry, cohen2019geometric}.

Despite the considerable insights achieved by these works, the approaches they employ make it difficult to understand the physical origin of topological protection. The use of transition matrices (or closely related and even more mathematically complex objects such as tilted current matrices) seems to suggest that topological protection is a property of the \emph{kinetics} of the system under consideration~\cite{murugan2017topologically, dasbiswas2018topological, tang2021topology}. Furthermore, the relationship between topological protection and nonequilibrium properties such as fluxes and thermodynamic driving forces remains obscure.

Here, we revisit topological protection in chemical systems and show that this phenomenon can be understood in terms of the theory of nonequilibrium thermodynamics of steady-states (NESS) in open systems~\cite{qian2005thermodynamics, qian2007phosphorylation, beard2008chemical}. By reformulating the physical description of these systems in terms of stoichiometric matrices and driving forces, we show that the origin of topological protection in these systems is primarily \emph{thermodynamic} rather than kinetic. Our formalism allows us to directly link the existence of topologically protected edge modes and edge currents to driving forces across elementary reactions that make up the system. We demonstrate these ideas by revisiting simple examples inspired by earlier works~\cite{knebel2020topological, tang2021topology}.

To make a closer connection to physical kinetics, we focus primarily on open chemical systems (e.g., cells) where a subset of chemical species (e.g., ATP and ADP) are maintained out-of-equilibrium and used to drive chemical reactions. However, most of these ideas naturally transfer to a more general setting based on nonequilibrium steady-states described by Master Equations~\cite{ge2010physical}.

To begin, we consider a simple chemical network composed of chemical species (indexed by $i$) with average concentrations $x_i$ that can participate in \emph{reversible} reactions (indexed by $\alpha$). Reversibility of the reactions is necessary for thermodynamic  consistency. The dynamics of such system are described by the kinetic equations
\be
{d c_i \over dt} = \sum_{\alpha} S_{i \alpha} J_\alpha +\bext_i,
\label{Eq:dynamics1}
\ee
where $J_\alpha$ is the net flux of reaction $\alpha$, $\bext_i$ is the rate at which species $i$ is produced and degraded by external sources or sinks (we will generally assume all $\bext_i=0$), and $S_{i \alpha}$ is the stoichiometric matrix which encodes the number of each species $i$ consumed (negative entries) or produced (positive entries) by reaction $\alpha$~\cite{qian2005thermodynamics, qian2007phosphorylation, beard2008chemical}. Since all reactions are reversible, we further divide the net flux into a forward flux $J_\alpha^+$ and a backward flux $J_\alpha^-$ with
${J_\alpha= J_\alpha^+ - J_\alpha^-}$.
These two fluxes can then be related to the concentrations via the law of mass action,
\be
J^\pm_\alpha = k^\pm_\alpha \prod_{i, S_{i \alpha} \lessgtr 0}  c_i^{\abs{S_{i\alpha}}},\label{Eq:massaction}
\ee
where $k^\pm_\alpha$ are the forward and backward rate constants of reaction $\alpha$.

As an example, consider a reaction $\gamma$ of the form
\begin{equation*}
 2 X_l \xrightleftharpoons[k_{\gamma}^-]{k_{\gamma}^+} 3 X_m +X_n
\end{equation*}
that consumes two molecules of $X_l$ and produces three molecules of $X_m$ and one molecule of $X_n$.
The associated entries in the stoichiometric matrix are ${S_{l \gamma}=-2}$, ${S_{m \gamma}=3}$, ${S_{n \gamma}=1}$ or zero otherwise with fluxes ${J_\gamma^+= k_{\gamma}^+ c_l^2}$ and ${J_\gamma^-=k_{\gamma}^-c_m^3 c_n}$. 

We now exploit some beautiful results from the theory of NESS to reexpress these kinetic equations in terms of thermodynamic driving forces. 
To begin, we write the ratio of the forward and backward fluxes, $J_\alpha^+$ and $J_\alpha^-$, in terms of the thermodynamic driving force $\Delta \mu_\alpha$ (chemical potential difference) using the NESS identity
\be
{J_\alpha^- \over J_\alpha^+}= e^{{\Delta \mu_\alpha \over k_BT }},
\label{Eq:currentpotential}
\ee 
where $k_B$ is Boltzmann's constant and $T$ is the temperature (see SI and Refs.~\onlinecite{qian2005thermodynamics, qian2007phosphorylation, beard2008chemical}). 

Next, we decompose $\Delta \mu_\alpha$ into three distinct terms:
\be
\Delta \mu_\alpha = \Delta \mu_\alpha^0 + \drive_\alpha + k_B T \sum_i S_{i \alpha} \ln{c_i},
\label{Eq:defchemicalpotential}
\ee
with $ \Delta \mu_\alpha^0 = -k_B T \sum_i S_{i \alpha} \ln{c_i^{eq}}$ the chemical potential difference at equilibrium (in the absence of coupling to any high-energy molecules such as ATP/ADP, GTP/GDP, etc.) and $\drive_\alpha$ the nonequilibrium free energy associated with driving reaction $\alpha$ by coupling to high-energy molecular processes such as phosphorylation. The last term is the usual concentration-dependent contribution to the chemical potential difference. In kinetic descriptions, coupling to the high-energy molecules $\drive_\alpha$ is often included \emph{implicitly} and for this reason, we can identify the first two terms in the driving force with the kinetic parameters $k_\alpha^\pm$ in Eq.~\eqref{Eq:massaction} for the law of mass action (see SI and Ref.~\onlinecite{ qian2007phosphorylation}),
\be
 \Delta \mu_\alpha^0 + \drive_\alpha= k_BT \ln{k_{\alpha}^- \over k_{\alpha}^+}.
\label{Eq:parametersdriverelation}
\ee

Using the NESS identity of Eq.~\eqref{Eq:currentpotential}, we then write the net flux ${J_\alpha=J_\alpha^+-J_\alpha^-}$ in terms of only the backward fluxes $J_\alpha^-$ and driving force $\Delta \mu_\alpha$ to get
\be
J_\alpha  =(e^{-{\Delta \mu_\alpha \over k_BT}}-1) J_\alpha^- \equiv \theta_\alpha J_\alpha^-,
\label{Eq:deftheta}
\ee
where the second equality defines the parameter $\theta_\alpha$.  This parameter, which we call the \emph{thermodynamic drive}, will play a central role in what follows, and for this reason, it is worth gaining some basic intuition for its meaning. When $\theta_\alpha>0$  the net reaction flux is in the forward direction. In contrast, when $\theta_\alpha<0$ the net flux is in the negative direction. When $\theta_\alpha =0$, there is no net current across reaction $\alpha$ and the reaction is locally at detailed balance. Consequently, all $\theta_\alpha$ are identically zero for a system at equilibrium. From these considerations, it is clear that $\theta_\alpha$ is a measure of the nonequilibrium drive associated with reaction $\alpha$.

One subtlety worth emphasizing is that $\theta_\alpha$ depends not only on the kinetic parameters via the driving potentials $\drive_\alpha$, but also on the steady-state concentrations $c_i$ [see Eqs.~\eqref{Eq:defchemicalpotential} and~\eqref{Eq:deftheta}]. For this reason, we cannot directly control $\theta_\alpha$ in our systems by changing kinetic parameters. However, it is still possible to choose $\drive_\alpha$ to prejudice $\theta_\alpha$ towards a particular direction. Physically, this corresponds to changing the amount of free energy a reaction can transduce when it is coupled to high-energy processes such as phosphorylation~\cite{hill2012free, qian2007phosphorylation}. 
Finally, we note that the choice of expressing Eq.~\eqref{Eq:deftheta} in terms of $J^-_\alpha$ or $J^+_\alpha$ is arbitrary (i.e. a choice of gauge).
Although each choice results in a slightly different definition of  $\theta_\alpha$, 
it will follow the same sign convention in either case, so this choice does not affect our analysis.

With this basic picture in mind, we write the steady-state condition for Eq.~\eqref{Eq:dynamics1} entirely in terms of negative fluxes,
\be
0=\sum_\alpha S_{i \alpha}^- J_\alpha^-,
\label{Eq:steady-state}
\ee
where we have set ${\bext_i=0}$ and defined the drive-dependent stoichiometric matrix ${S_{i \alpha}^- \equiv S_{i \alpha}\theta_\alpha}$.
We show below that $S_{i \alpha}^-$ encodes all the information necessary to identify topological phase transitions.
To motivate this, we note that $S_{i \alpha}^-$ plays a similar role to the compatibility matrix in topological mechanics, but for a system with directed interactions.
We also emphasize that $S_{i \alpha}^-$ is a purely thermodynamic quantity; it only depends on the topology of the chemical network and the thermodynamic drives of each reaction.
Effectively, Eq.~\eqref{Eq:steady-state} is analogous to Kirchhoff's node rule for chemical networks expressed purely in terms of backward fluxes.


\begin{figure}[t!]
\includegraphics[width=0.95\columnwidth]{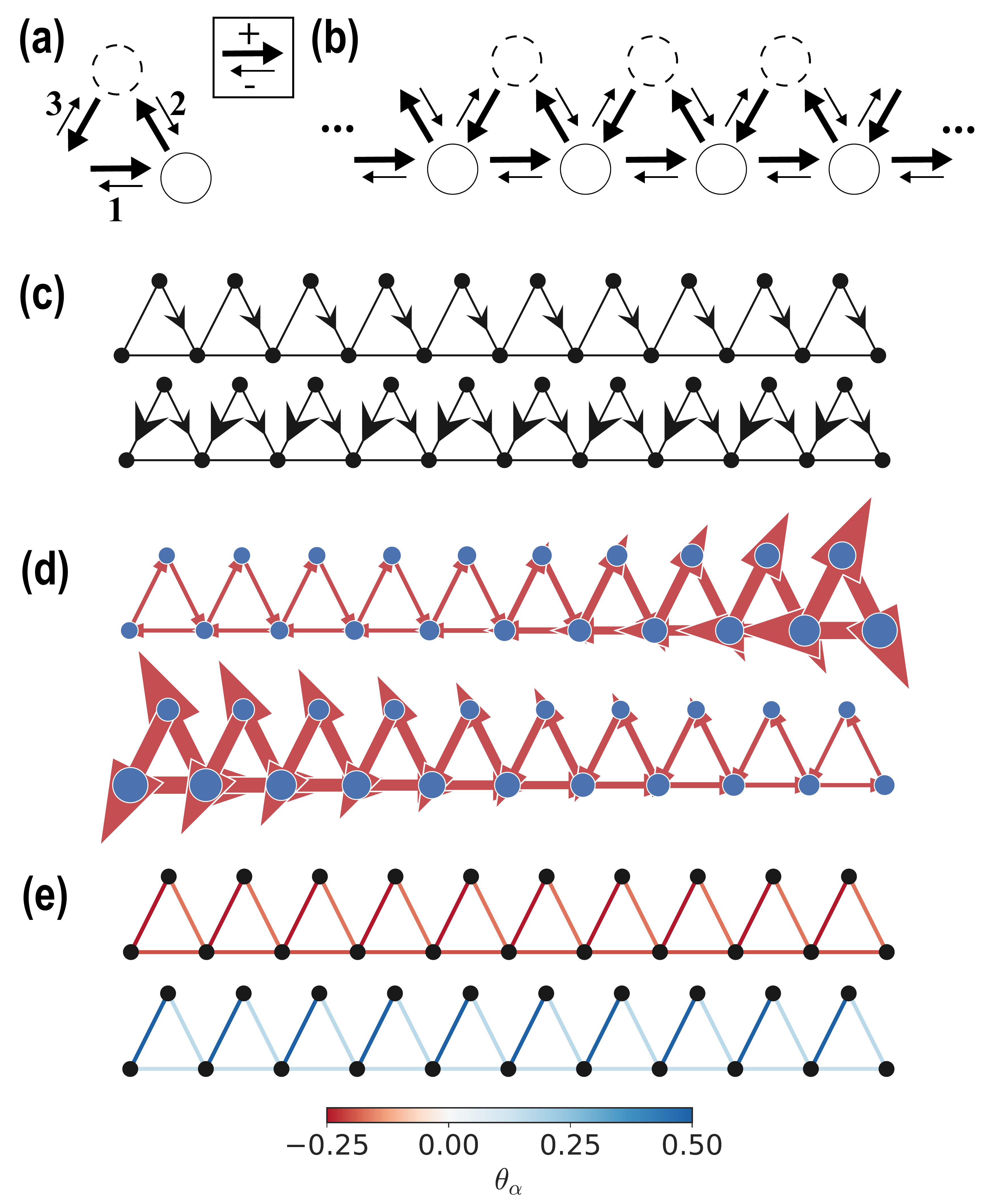} 
\caption{Topologically-protected edge modes in a network of chemical reactions on a triangular lattice. (a) The primitive cell is composed of two types of chemical species and three types of reversible reactions. Forward/backward directions correspond to large/small arrows. (b) The full chemical network is formed by repeating the basic primitive cell. 
(c) The driving chemical potentials $\Delta \mu^0_\alpha + \drive_\alpha$ for two different systems with $N=10$ lattice cells. The size and direction of the arrows indicate the magnitude and direction of chemical potential drops across each edge.
The kinetic parameters are  $k_1^\pm=k_2^+=k_3^+=1$, $k_2^-=2$, and $k_3^-=1$ ($0.25$) for the upper (lower) configuration, corresponding to $\Delta \mu^0_1 + \drive_1 = 0$, $\Delta \mu^0_2 + \drive_2 = k_BT\log 2$, and $\Delta \mu^0_3 + \drive_3 = 0$ ($-2k_BT\log 2$).
(d) Numerical simulations of the systems in (c) showing the existence of left- and right-localized edge modes. 
Magnitudes of net nonequilibrium fluxes $J_\alpha$ are indicated by thickness of arrows and steady-state concentrations by sizes of nodes. 
(e) Thermodynamic drives $\theta_\alpha$ of the two systems indicated by edge color with $\theta_\alpha > 0$ ($<0$) corresponding to a drive in the forward (backward) direction.
}
\label{Fig:fig1}
\end{figure}

To demonstrate topological protection within this framework,
we start by analyzing a simple triangular lattice network inspired by a recent paper on topological phase transitions in a game-theoretic model of a coupled rock-paper-scissor system~\cite{knebel2020topological}. 
As shown in Fig.~\ref{Fig:fig1}(a), each unit cell consists of three chemical reactions (edges) that interconvert two different chemical species (nodes).
Repeating this unit $N$ times [Fig.~\ref{Fig:fig1}(b)] and closing the system with an additional boundary node forms a chain of $2N+1$ chemical species coupled by $3N$ reactions.
All reactions conserve particle number with each reaction $\alpha$ between species $i$ and $j$ simply converting one molecule of species $X_i$ to one molecule of species $X_j$ and vice versa,
\begin{equation*}
X_i \xrightleftharpoons[k_{\alpha}^-]{k_{\alpha}^+} X_j.
\end{equation*}
Furthermore, we assume that each unit cell has the same three types of reactions, $A=1,2,3$ with kinetic parameters $k_A^{\pm}$ and thermodynamic drive $\theta_A$ [Eqs.~\eqref{Eq:defchemicalpotential} and ~\eqref{Eq:deftheta}].

Using standard arguments, we write the full stoichiometric matrix in terms of the $2$-by-$3$ stoichiometric matrix of the primitive lattice cell in Fourier space (see SI),
\be
\hat{S}(k)=\mqty[1-e^{ik} & -1 & e^{ik} \\
0 & 1 &-1],
\ee
where rows correspond to chemical species and columns to fluxes. We interpret  $e^{ik}$ as a ``shift operator" that translates the flux by a unit cell to the right, allowing for a simple physical interpretation of the elements of $\hat{S}(k)$ as subtracting and adding fluxes in adjacent unit cells. The resulting Fourier transform of $S_{i \alpha}^-$ [Eq.~\eqref{Eq:steady-state}] is simply the $2 \times 3$ matrix
\be
\hat{S}^-(k)=\mqty[(1-e^{ik})\theta_1 & -\theta_2 & e^{ik}\theta_3\\
0 & \theta_2 &-\theta_3],
\ee
where once again rows correspond to chemical species, but columns now correspond to backward fluxes ($J_\alpha^-$).

To identify topological phase transitions, we analyze the spectrum of $\hat{S}^-(k)$ or equivalently, the Hermitian matrix ${\hat{H}(k) = \hat{S}^-(k) \hat{S}^-(k)^\dag}$ whose eigenvalues are the squares of the singular values of $\hat{S}^-(k)$,
\bea
\hat{H}(k)&=&\mqty[
2(1-\cos{k})\theta_1^2+\theta_2^2+\theta_3^2 &\, -\theta_2^2-\theta_3^2 e^{ik} \\
-\theta_2^2-\theta_3^2 e^{-ik} &\, \theta_2^2+\theta_3^2]. 
\eea
We also further restrict our analytics to the special case $\theta_1=0$ where reactions of type $A=1$ are in detailed balance (attainable in an infinite system by setting ${\Delta \mu^0_1 + \drive_1 = 0}$, see SI).
In this case, we can write ${\hat{H}(k)=a_0 I + \vec{a} \vec{\sigma}}$
where $I$ is the identity matrix and $\vec{\sigma}$ is the set of Pauli matrices with
${a_0=\theta_2^2 + \theta_3^2}$ and ${\vec{a} = (-\theta_2^2-\theta_3^2 \cos{k}, -\theta_3^2 \sin{k}, 0)}$ (see SI). In terms of $\vec{a}$ and $a_0$, the spectrum is 
\be
\lambda_\pm = a_0 \pm \sqrt{\vec{a}\cdot \vec{a}}= a_0 \pm\sqrt{\theta_2^4 + \theta_3^4+2\theta_2^2 \theta_3^2 \cos{k}},
\ee 
which becomes degenerate at $k=\pi$ when $\theta_2^2=\theta_3^2$. 

To see that this corresponds to a topological phase transition, we follow standard arguments and plot the $x$- and $y$-components of $\vec{a}$ as $k$ varies from $0$ to $2\pi$ (see SI Fig.~\ref{SIfig:fig1}). For all parameters, the resulting curve is a circle of radius $\theta_3^2$ centered at the point $(-\theta_2^2, 0)$ in the $x$-$y$ plane. When $\theta_3^2 < \theta_2^2$, the circle does not enclose the origin, whereas if $\theta_3^2 > \theta_2^2$, this curve always contains the origin with the transition occurring exactly when $\theta_3^2 = \theta_2^2$ and the gap closes. Since the curve $(a_x(k), a_y(k))$ can be viewed as a map from $S^1$ to $S^1$ (recall $k$ is periodic with period $2 \pi$), this corresponds to a change in the winding number and hence defines a $\mathbb{Z}_2$ topological invariant~\cite{nakahara2018geometry}. Alternatively, one can show the existence of a Zak phase by interpreting $\hat{H}(k)$ as a quantum Hamiltonian and making use of well-established classifications of topological phases~\cite{chiu2016classification}. As in the original game theoretic example considered in Ref.~\onlinecite{knebel2020topological}, our analytics suggest that the topological transition is controlled by a single parameter related to the strength of reactions of types 2 and 3. However, in our system, this transition depends on the kinetic rate constants only through the ratio of nonequilibrium thermodynamic drives.
 
To test these predictions, we ran numerical simulations on a triangular lattice with closed boundary conditions. We numerically integrated the kinetic equations of Eq.~\eqref{Eq:dynamics1} assuming mass action [Eq.~\eqref{Eq:massaction}] to calculate steady-state concentrations $c_i$ and fluxes $J_\alpha$ (see SI for details). Fig.~\ref{Fig:fig1}(c)-(e) shows results from simulations where we chose $k_1^\pm = k_2^+=k_3^+=1$, $k_2^-=2$ and varied a single parameter, $k_3^-$. Thermodynamically, this is equivalent to controlling the thermodynamic driving force of reaction 3, $\Delta \mu_3^0+ \drive_3$. 
Using Eq.~\eqref{Eq:deftheta}, we estimated $\theta_\alpha$ for each edge and separately computed the mean-squared average  over all reactions of the same type,  $\<\theta_A^2\>$.
We find that the system robustly supports two types of edge states localized to different ends of the lattice. 
Consistent with our analytics, when $\<\theta_3^2\> \gg \<\theta_2^2\>$, the systems exhibits right localized edge states and when $\<\theta_3^2\> \ll \<\theta_2^2\>$ the system has left localized states. However, our numerics are inadequate to probe the transition region since all reactions in our system are reversible resulting in large finite-size effects. 


\begin{figure}[t!]
\includegraphics[width=0.95\columnwidth]{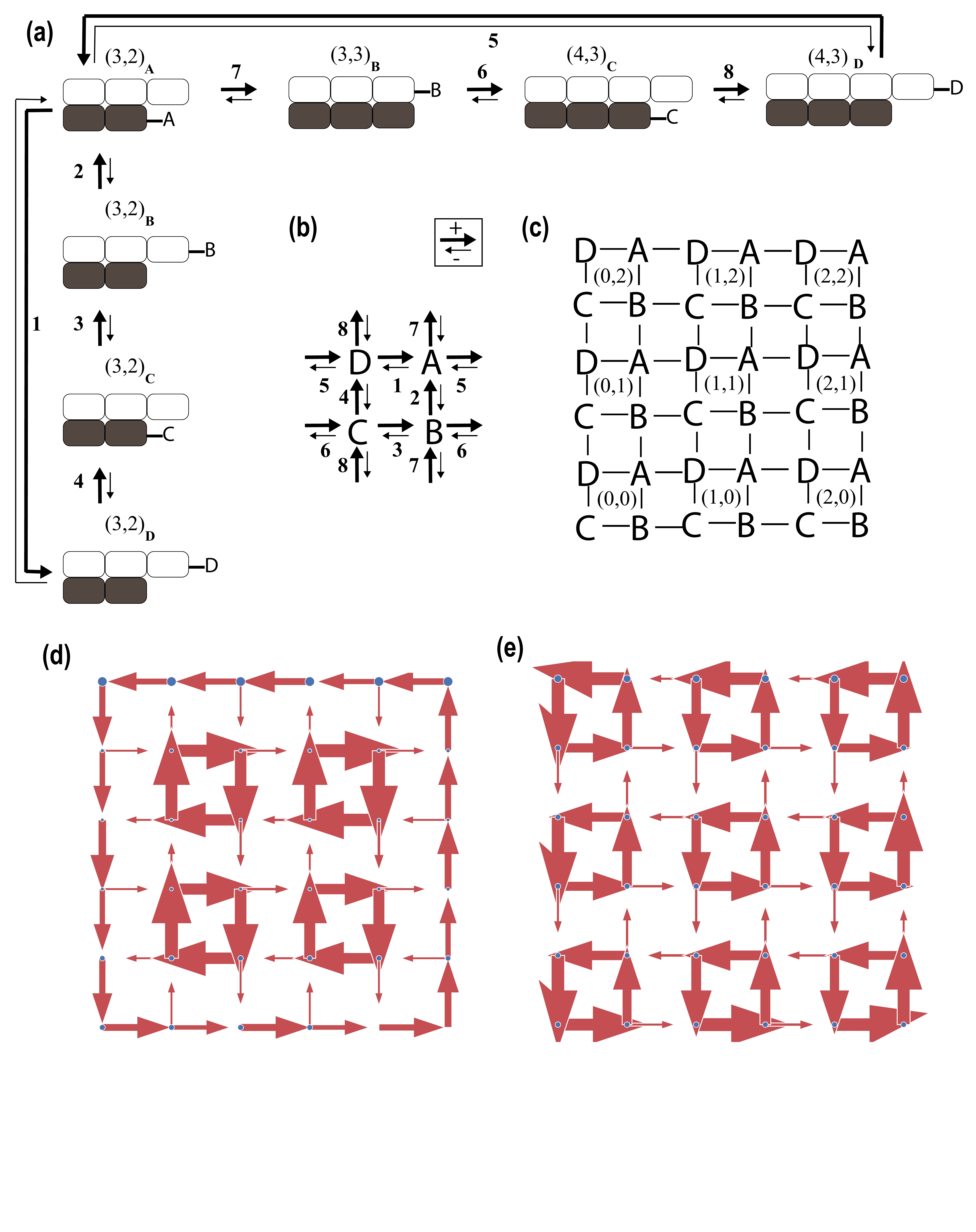} 
\caption{Topologically-protected edge current in a simple model of biopolymer self-assembly where the state of the biopolymer is parameterized by the number of monomers of two different types, along with one of four possible internal states.
(a) Schematic of eight possible elementary reactions where the internal reactions (1-4) transition between the internal states of the biopolymer and the external reactions (5-8) add or remove monomers. (b) Primitive lattice cell with four chemical species corresponding to the four internal states (nodes) and eight fluxes (edges). Forward/backward directions correspond to large/small arrowheads. (c) A chemical reaction network with $N_x=N_y=3$. (d) 
Stochastic simulations using the Gillespie algorithm showing a typical nonequilibrium steady-state flux configurations with a topologically protected chiral edge current ($k_a^+=9k_a^-$) and (e) without an edge current ($k_a^+=k_a^-/3$). Magnitudes of nonequilibrium fluxes are indicated by thickness of arrows and steady-state concentrations by sizes of nodes. For both simulations: $k_a^-= k_b^-=0.01$, $k_b^+=3k^-$.}
\label{Fig:fig2}
\end{figure}

Next, we analyze an elegant toy model for self-assembly of biopolymers shown Fig.~\ref{Fig:fig2}, previously introduced in Ref.~\onlinecite{tang2021topology}.  This model was previously shown to support topological edge currents using an approach based on the Master Equation and techniques inspired by non-Hermitian quantum systems~\cite{shen2018topological, kawabata2019symmetry, lieu2018topological}. In this model, a biopolymer molecule can expand or contract by adding or losing two different types of monomers, $X$ and $Y$. At a given time, the biopolymer is described by an ordered pair of integers $(x,y)$ representing 
the number of monomers of type $X$ and $Y$ currently incorporated into the molecule, along with a parameter $z=A,B,C,D$ that can take on one of four values corresponding to 
four different ``internal states'' of the biopolymer. 
Depending on the internal state, the 
biopolymer can add or remove a monomer of $X$ and $Y$ [reactions 5-8 in Fig. 
\ref{Fig:fig2}(a)]. The biopolymer molecule can also cyclically switch between internal states through 
four additional reactions [labeled 1-4 in Fig.~\ref{Fig:fig2}(a)]. Each of these eight reactions $
\alpha$ has two associated kinetic parameters $\{k_\alpha^\pm\}$ and a thermodynamic driving force $\theta_\alpha$. We also assume that the biopolymer can 
incorporate at most $N_x$ of the $X$ monomers and $N_y$ of the $Y$ monomers so that  
$0\le x\le N_x$ and $0\le y \le N_y$. As shown in Fig.~\ref{Fig:fig2}(c),  this chemical reaction 
network naturally lives on a two-dimensional lattice in which each node represents an internal state and each edge represents a chemical reaction. 
In Ref.~\onlinecite{tang2021topology}, the analytics were largely confined to the ``chiral'' case in which all reactions were strictly irreversible.
Here, we instead consider a more general case where all reactions are reversible.

As shown in Fig.~\ref{Fig:fig2}(b), the elementary lattice cell for this chemical reaction network is composed of four chemical 
species (corresponding to each of the four possible internal states of the biopolymer at a 
fixed monomer composition) and eight elementary reactions that add or remove 
monomers or change the internal state. By 
using a procedure analogous to that used for the triangular network, it is straightforward to 
calculate the Fourier transform for the drive-dependent stoichiometric matrix $\hat{S}^-(\vec{k})
$, where $\vec{k}=(k_x,k_y)$ includes the momentum in both the horizontal and vertical directions (see SI). To look for topological phase transitions, we once again analyze the spectrum of $\hat{H}(\vec{k})=\hat{S}^-(\vec{k})\hat{S}^-(\vec{k})^\dag$ and ask where an eigenvalue gap closes.
For simplicity, we restrict our analysis to the case where the thermodynamic drives associated with each of the four internal reactions are identical with $\theta_\alpha = \theta_a$ for $\alpha=1,2,3,4$ and similarly the external reactions are identical with $\theta_\alpha=\theta_b$ for $\alpha=5,6,7,8$  (attainable by separately setting $\Delta \mu^0_\alpha + \drive_\alpha$ equal for the internal and external reactions, see SI).
We find that the gap in the 
spectrum closes when the internal and external driving forces are equal with $\theta_a^2=\theta_b^2$. 
This closing of the gap is also associated with a change in a value 
of a  $\mathbb{Z}_2$ topological invariant (i.e., a nontrivial Zaks phase) (see SI). Furthermore, in analogy with the integer Hall effect, we expect to see a chiral edge current emerge in one of these phases (as discussed in Ref.~\onlinecite{tang2021topology}).

To check this prediction numerically, we performed stochastic simulations using the Gillespie algorithm \cite{gillespie1976general}. In our simulations, we set $k_a^-=k_b^-=0.1$, $k_b^+=0.27$ and varied the ratio of $\theta_a$ and $\theta_b$ by varying $k_a^+$  (or equivalently, varying $\Delta \mu_a^0 + \drive_a$). As predicted, we found our system robustly supports chiral edge currents over a large range of parameters. Typical nonequilibrium steady-state configurations observed in our simulations with and without edge currents are shown in Figs.~\ref{Fig:fig2}(d) and (e).

In conclusion, we have revisited the idea of topological protection in open chemical systems and shown that the origin of topological protection can be understood directly in terms of nonequilibrium thermodynamic quantities. In our analysis, thermodynamic driving forces play a fundamental role: they control the sign and magnitude of nonequilibrium fluxes, as well as the stoichiometric matrix which encodes the topology/geometry of interactions. This is consistent with the well-established result from the theory of nonequilibrium thermodynamics that steady-states can be understood by decomposing fluxes into cycles whose magnitude is controlled by chemical potential differences/theromodynamic forces across the cycle~\cite{hill2012free}. Here, we have limited our analysis to some simple systems with just a few types of reactions. We expect that for more complicated systems (e.g., the case where all eight reactions in the biopolymer example are distinct) systems will likely exhibit a much richer phase diagram with multiple topological phase transitions. Our analysis also raises many new questions. Previous work explicitly makes use of the non-Hermitian nature of the transition matrices defining the Markov process~\cite{tang2021topology}. Here, we instead analyzed the spectrum of a  Hermitian matrix $\hat{H}(k)$. Given that non-Hermitian systems exhibit unique topological properties~\cite{shen2018topological, kawabata2019symmetry, lieu2018topological}, much more work is needed to reconcile the thermodynamic picture presented here with one based on kinetics. Doing so will likely lead to considerable physical insight and illuminate deep mathematical structures relating thermodynamics, stochastic systems, and topology.

\section*{Acknowledgments}

We thank Evelyn Tang for helpful discussions and Wenping Cui for comments on the manuscript. This work was supported by NIH R35GM119461  to PM.

\bibliography{refsmain.bib}

\onecolumngrid
\newpage

\appendix

\renewcommand{\thefigure}{\hbAppendixPrefix\arabic{figure}}
\setcounter{figure}{0}

\section{Review of basic nonequilibrium thermodynamic steady-states}

We start by reviewing some basic ideas on how to describe the nonequilibrium steady-states (NESS) of open chemical systems. In doing so, we largely follow the formalism
developed by H. Qian and D. Beard in \cite{beard2008chemical, qian2005thermodynamics, qian2007phosphorylation}. 

The fundamental objects in our formalism for describing a chemical reaction network are the concentrations of chemical species $c_i$ that can undergo different \emph{reversible} chemical reactions, labeled by $\alpha$. Reversibility is necessary in order to have a consistent thermodynamic description of the system.

\subsection{Fluxes in Open Systems}
Each reaction $\alpha$ is characterized by a net flux $J_\alpha$, that can be further decomposed into a forward flux $J_\alpha^+$ and a backward flux $J_\alpha^-$ with
\be
J_\alpha=J_\alpha^+ - J_\alpha^-.
\label{SIeq:def_J}
\ee
 In equilibrium, we have detailed balance so that $J_\alpha^+=J_\alpha^-=J_\alpha^{ \pm eq}$ and there is no net current $J_\alpha=0$. However, first emphasized by Hill \cite{hill2012free}, and then Qian and colleagues \cite{beard2008chemical, qian2005thermodynamics, qian2007phosphorylation}, in open chemical systems like cells where the concentration of some chemical species are held fixed (e.g., ATP and ADP concentrations inside a cell), reactions can be driven by consuming energy. In this case, we know in general that $J_\alpha \neq 0$ and we can define a thermodynamic driving force/chemical potential $\Delta \mu_\alpha$ associated with each reaction,
\be
\Delta \mu_\alpha = k_B T \ln{J_\alpha^- \over J_\alpha^+}= k_B T \ln{\left(1-{J_\alpha \over J_\alpha^+}\right)}.
\label{SIeq:def_deltamu}
\ee
We can combine Eqs.~\eqref{SIeq:def_J} and \eqref{SIeq:def_deltamu} to write
\be
J_\alpha=J_\alpha^+ - J_\alpha^-= (e^{-{\Delta \mu_\alpha \over k_B T}} -1)J_\alpha^-.
\label{SIJchem}
\ee

Near equilibrium, $J_\alpha \ll J_\alpha^{ \pm eq}$ so that Eq.~\eqref{SIeq:def_deltamu} reduces to the usual linear relationship between drives and fluxes for linear response,
\be
\Delta \mu_\alpha \approx -{k_B T \over J_\alpha^{ \pm eq} }J.
\ee
Finally, we note that we can define the entropy production rate $epr$ as
\be
epr= -\sum_\alpha \Delta \mu_\alpha J_\alpha \ge 0,
\ee
where the inequality follows from substituting Eqs.~\eqref{SIeq:def_J} and \eqref{SIeq:def_deltamu} into the expression above. 

Eq.~\eqref{SIeq:def_deltamu} relating $\Delta \mu_\alpha$ to the ratio of the forward and backward flux will play a central role in what follows.

\subsection{Stoichiometric Matrix}

The second important ingredient needed to describe NESS in our formalism 
is the stoichiometric matrix $S_{i \alpha}$, where $i$ runs over chemical
species and $\alpha$ runs possible reactions. As the name suggests,
$S_{i \alpha}$  encodes the stoichiometry of the reactions. In particular, $S_{i \alpha}$ encodes how many molecules $i$ are consumed (negative entries) or produced by reaction $\alpha$ (positive entries). If a reaction $\alpha$ consumes two molecules of species $l$ to produces a molecule of species $m$ and a molecule of species $n$ then $S_{l \alpha}=-2$, $S_{m \alpha}=1$ and $S_{n \alpha}=1$ with $S_{i \alpha}=0$ if $i \neq l,m,n$.

By definition, the dynamics of species $i$ are described  by the differential equation
\be
{d c_i \over dt} = \sum_{i \alpha} S_{i \alpha} J_\alpha + \bext_i,
\ee
where $\bext_i$ is the rate at which $c_i$ is produced or depleted by external sources or sinks (i.e., produced/depleted by reactions not included in the network or through exchange with the environment). For a NESS, we can set the left side to zero and we have 
\be
0 = \sum_{i \alpha} S_{i \alpha} J_\alpha + \bext_i
\ee
For chemical species that are are not externally replenished or depleted, this reduces to the familiar
Flux Balance equation,
\be
0 = \sum_{ \alpha} S_{i \alpha} J_\alpha.
\ee

\subsection{Chemical potential differences for reactions}

We can also associate each reaction $\alpha$ with a chemical potential difference $\Delta \mu_\alpha$ which can be decomposed into three parts:
\be
\Delta \mu_\alpha = \Delta \mu_\alpha^{0}+\drive_\alpha + k_B T\sum_i S_{i \alpha} \ln{c_i},
\ee
where $\Delta \mu_\alpha^{0}$ is just the equilibrium chemical potential difference characterizing the reactions in the \emph{absence} of nonequilibrium driving (i.e., the difference in the \emph{ absence} of any coupling to high-energy molecules maintained out of equilibrium by cells such as ATP/ADP, GTP/GDP, etc.), the term $\drive_\alpha$ represents the transduction of free energy by coupling the reaction to a to high-energy molecular processes such as phosphorylation~\cite{hill2012free}, and the last term encodes the usual dependence of the free energy differences on the concentrations of reactants and products. 

It is worth better understanding this decomposition in greater detail. To do so, first consider the two following simple chemical reactions:
\bea
A &\xrightleftharpoons[k_0^-]{k_0^+}& B  \label{SIeq:nodrive} \\
A + \mathrm{ATP} &\xrightleftharpoons[k^-]{k^+}& B +  \mathrm{ADP}+ \mathrm{P} \label{SIeq:driven}
\eea

For Eq.~\eqref{SIeq:nodrive} without a driving force, we know that $\drive_\alpha=0$ and $\Delta \mu= \Delta \mu^0+k_BT \ln{[B] \over [A]}$. Since at equilibrium, we know that $\Delta \mu=0$ and detailed balance is satisfied, $k_0^-[B]_{eq}=k_0^+ [A]_{eq}$. For this reason, we conclude that we must have the usual relation between kinetic constants and thermodynamic chemical potentials,
\be
\Delta \mu^0=k_B T \ln{k_0^- \over k_0^+}.
\ee

Now consider the second driven reaction in Eq.~\eqref{SIeq:driven}. In this case, we have by definition that ~\cite{qian2007phosphorylation}
\be
\Delta \mu= \Delta \mu^0- k_B T \ln{[ATP] \over [ADP][P]}+ k_BT \ln{[B] \over [A]}.
\ee
Thus, we can identify 
\be
\drive = - k_B T \ln{[ATP] \over [ADP][P]}.
\ee
As emphasized by Qian, Beard, and colleagues, we can actually just implicitly treat the high-energy phosphorylation reactions be defining effective forward and backward rates
$k_+=k_0^+ [ATP]$ and $k_-=k_0^- [ADP][P]$. In terms, of these effective rates for Eq.~~\eqref{SIeq:driven}, we have 
\be
\Delta \mu =k_B T \ln{k_- \over k_+}+ k_BT \ln{[B] \over [A]}.
\ee
Notice that this is just the special case of our general formula in Eq.~\eqref{SIeq:def_deltamu} that 
\be
\Delta \mu_\alpha =k_B T \ln{J^- \over J^+}.
\ee

More generally, we see that a very similar argument gives
\be
\Delta \mu_\alpha = k_B T \ln{k_{\alpha}^- \over k_{\alpha}^+}+ \sum_{i} S_{i \alpha}\ln{c_i},
\ee
where we have defined ``effective rate constants'' $k_{\alpha}^-$ and $k_{\alpha}^+$ that implicitly include any high-energy molecules involved in the reactions. Furthermore, we know
that we can identify the driving force with the chemical potential difference,
\be
\drive_\alpha = k_B T \ln{k_{\alpha}^- \over k_{\alpha}^+}-k_B T \ln{k_{\alpha 0}^- \over k_{\alpha 0}^+},
\ee
or equivalently,
\be
\drive_\alpha+ \Delta \mu_\alpha^0=k_B T \ln{k_{\alpha}^- \over k_{\alpha}^+}.
\label{SIEq:k_+k_-}
\ee


\section{Analytics: Edge modes in one-dimensional triangular lattice}

We now give a detailed analysis of the triangular lattice discussed in the main text.
\subsection{Defining generalized stoichiometric matrix}
Consider a periodic triangle network of the type shown in Fig.~\ref{Fig:fig1}. The primitive
lattice for this consists of three fluxes and two nodes which are repeated $N$ times. For a such a lattice, we can consider the stoichiometric matrix in Fourier space. This matrix is given by
\be
\hat{S}(k)=\mqty[
1-e^{ik} & -1 & e^{ik} \\
0 & 1 &-1],
 \label{SI:defhatS}
  \ee
where without loss of generality, we have set the lattice spacing to be one so that $e^{ik}$ represents a shift by a single lattice spacing. In Fourier space, the steady-state condition
\be
\sum_{\alpha} S_{i \alpha}J_\alpha=0
\label{SI:steady-state-eq}
\ee
becomes 
\be
\hat{S}(k) \hat{J}(k)=0,
\ee
where $\hat{J}(k)$ is just the usual Fourier transform of the flux distributions,
\be
\hat{J}_A(k)= \sum_{\alpha=1}^N e^{ik j} J_{A,\alpha},
\ee
where $A$ indicates the reaction type ($A=1,2,3$) and $\alpha$ ranges over the unit cells.

To understand the topological phase transitions we must work with slightly different objects.
Instead of writing our dynamics in terms of the \emph{net flux} $J_\alpha$, it is useful to write the flux in terms of the backward flux $J_\alpha^-$ (or alternatively, in terms of the forward flux $J_\alpha^+$. 
Substituting Eq.~\eqref{SIJchem} into Eq.~\eqref{SI:steady-state-eq} yields the modified steady state
equation
\be
\sum_{\alpha} S_{i \alpha}(e^{-{\Delta \mu_\alpha \over k_B T}} -1)J_\alpha^- = 0 .
\ee
Motivated by this equation, we define the parameter
\be
\theta_\alpha \equiv e^{-{\Delta \mu_\alpha \over k_B T}} -1,
\ee
which measures the direction and magnitude of the flux across reaction $\alpha$ and
the chemical potential-dependent stochiometric matrix for the negative fluxes 
\be
S_{i \alpha}^-\equiv S_{i \alpha} \theta_\alpha = S_{i \alpha}(e^{-{\Delta \mu_\alpha \over k_B T}} -1)
\label{SI:defSminus}.
\ee
For the the periodic triangular lattice [see Eq.~\eqref{SI:defhatS}], we can take the Fourier transform of $S_{i \alpha}^-$ to get 
\be
\hat{S}^-(k)= \mqty[(1-e^{ik}) \theta_1 & -\theta_2 & e^{ik} \theta_3\\
0 & \theta_2 &-\theta_3],
 \label{SI:defhatStriangle}
  \ee
 where $\theta_A = e^{-{\Delta \mu_\alpha \over k_B T}} -1$ is the drive associated with the three reactions $A=1,2,3$ associated with the three fluxes in the elementary lattices shown in Fig.~\ref{Fig:fig1}.
 
 \subsection{Spectrum and Topology} 
 
We now show that we can associate a $Z_2$ topological invariant with the triangular lattice by carefully analyzing the spectrum associated with $\hat{S}^-(k)$ as a function of $\theta_1$, $\theta_2$, and $\theta_3$ (or equivalently, the driving forces $\Delta \mu_1$, $\Delta \mu_2$, and $\Delta \mu_3$ associated with the three reactions).
 
 To do so, we can look at the singular values associated with $\hat{S}^-(k)$ or equivalently, the eigenvalues associated with the symmetric $2 \times 2$ matrix
 \be
\hat{H}(k)= \hat{S}^-(k) \hat{S}^-(k)^\dagger=\mqty[
2(1-\cos{k})\theta_1^2+\theta_2^2+\theta_3^2 &\, -\theta_2^2-\theta_3^2 e^{ik} \\
-\theta_2^2-\theta_3^2 e^{-ik} &\,\theta_2^2+\theta_3^2 ]
 \ee
 It is also helpful to decompose this Hamiltonian into a diagonal $k$-independent term,
 \be
 H_0=\mqty[\theta_2^2+\theta_3^2 & 0 \\
0 &\,\theta_2^2+\theta_3^2 ],
 \ee
 and a non-trivial $k$-dependent part ,
 \be
 \hat{H}_1(k)=\mqty[2(1-\cos{k})\theta_1^2 &\, -\theta_2^2-\theta_3^2 e^{ik} \\
-\theta_2^2-\theta_3^2 e^{-ik} &\, 0 ]
 \label{SI:defH1triangle}
 \ee
 so that
 \be
 \hat{H}(k)= H_0 + \hat{H}_1(k).
 \ee

To identify the topological phase transition,we rewrite $\hat{H}_1(k)$ in terms of the $2\times 2$ identity matrix $I$ as well as the three Pauli matrices $\sigma_x$, $\sigma_y$, and 
$\sigma_z$, allowing us to connect to standard arguments ~\cite{nakahara2018geometry}, 
 \be
 \hat{H}_1(k)= \theta_1^2(1-\cos{k})I -(\theta_2^2+\theta_3^2 \cos{k})\sigma_x
 -\theta_3^2 \sin{k} \sigma_y + \theta_1^2 (1-\cos{k})\sigma_z.
 \label{SI:defHtriangle}
 \ee
In particular, for a Hamiltonian of the form $H=a_0+\vec{a}\cdot \vec{\sigma}$, the two eigenvalues are given by
 \be
 \lambda_\pm = a_0 \pm \sqrt{\vec{a}\cdot \vec{a}}.
 \ee
 A $Z_2$ topological transition can occur when the gap between these two eigenvalues closes for any $k$. 
 
We now focus on the special case where $\Delta \mu_1=0$, or equivalently $\theta_1=0$ for which the topological transition occurs. This corresponds to the physical situation where reaction 1 is in detailed balance. 
This scenario can be attained by setting $\Delta \mu^0_1 + \drive_1 = 0$ and noting that the concentrations of the chemical species connected by type 1 reactions must be uniform due to translational symmetry.
As a result, $\Delta \mu_1$ and $\theta_1$ are guaranteed to be zero.
For this choice of parameters, $\hat{H}_1(k)=\vec{a}\cdot \vec{\sigma}$ with
 \be
 \vec{a}=(-(\theta_2^2+\theta_3^2 \cos{k}),  -\theta_3^2 \sin{k}, 0).
 \ee
 Notice that 
 \be
 \vec{a}\cdot \vec{a}= \theta_2^4+2\theta_2^2 \theta_3^2 \cos{k}+ \theta_3^4 
 \ee
 so that  for $k=\pi$ and $\theta_2^2=\theta_3^2$,  $\vec{a}\cdot \vec{a}=0$ and hence the gap in the spectrum closes. 
 
\begin{figure}[h!]
\includegraphics[width=0.95\columnwidth]{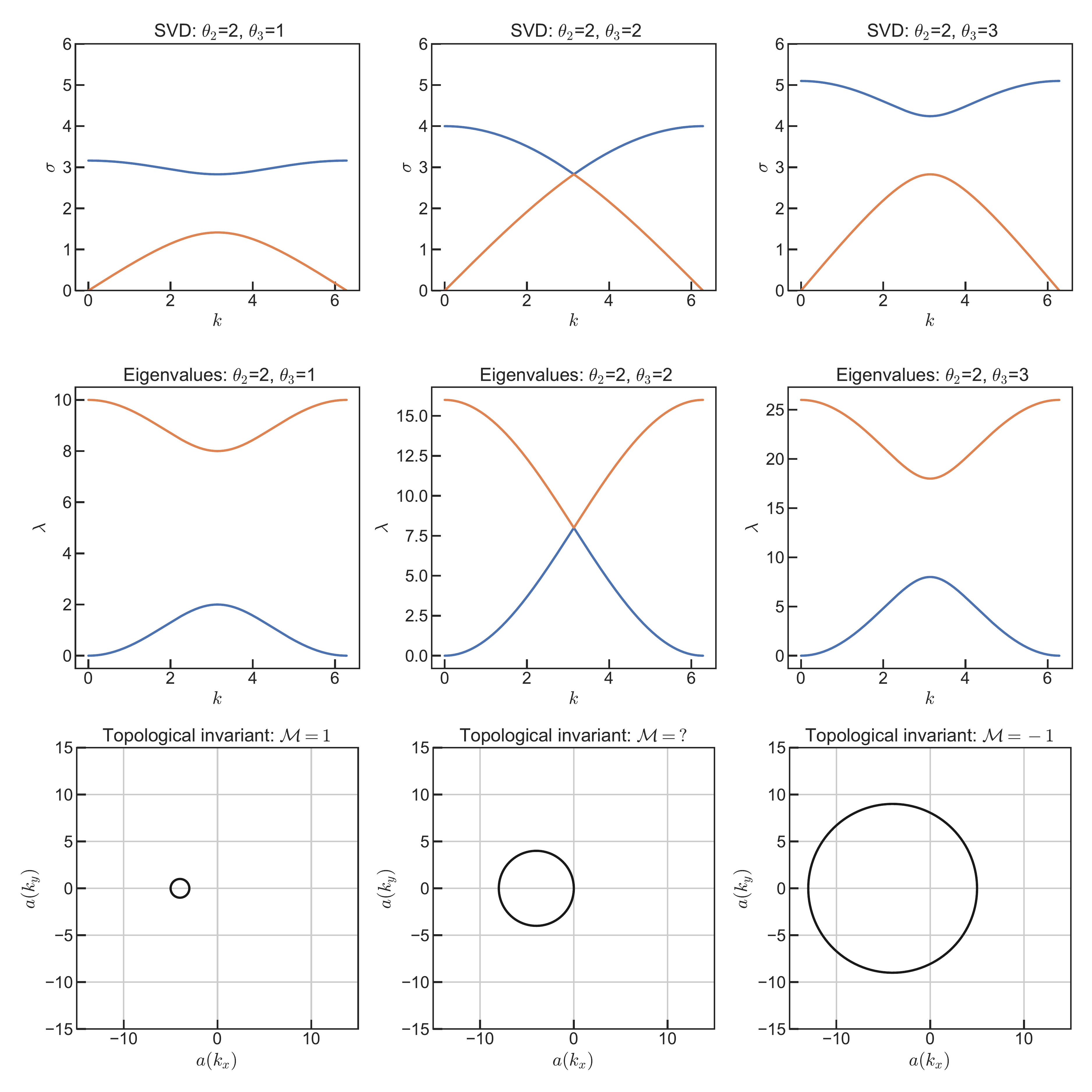} 
\caption{Analytics for triangular lattice. (Top) Singular value spectrum of $\hat{S}^-(k)$. (Middle) Eigenvalue spectrum of ${\hat{H}(k)=\hat{S}^-(k)\hat{S}^-(k)^\dag}$. (Bottom) Topological classification using map from $\hat{H}_1(k)=a_0+\vec{a}\cdot \vec{\sigma}$ for $k=[0,2\pi]$  to $(\sigma_x, \sigma_y)$. Curve is $(a_x(\vec{k}), a_y(\vec{k}))$. Topological invariant measures whether origin is enclosed.}
\label{SIfig:fig1}
\end{figure}

 To show that this is associated with a $\mathbb{Z}_2$ topological invariant,  we can plot the two-dimensional closed curve defined by the coefficients of $\sigma_x$ and $\sigma_y$, namely  $(a_x(k), a_y(k))$ where $k$ varies
 from $0$ to $2\pi$. A $\mathbb{Z}_2$ topological transition occurs at the values of $\theta_2$ and $\theta_3$ where this closed curve transitions from enclosing the origin to not enclosing the origin when the gap also closes. For $\hat{H}_1(k)$ in Eq.~\eqref{SI:defH1triangle}, this closed curve is given by
 \be
 (a_x(k), a_y(k))= (-\theta_2^2-\theta_3^2 \cos{k}, -\theta_3^2 \sin{k}).
 \ee
 These are just the coordinates for a circle with radius $\theta_3^2$ centered at the point $(0, -\theta_2^2)$. Thus, we see that there is a topological phase transition when $\theta_2^2=\theta_3^2$. For $\theta_2^2 < \theta_3^2$, the curve does not enclose the origin, whereas for $\theta_2^2 > \theta_3^2$ it does enclose the origin. This choice of parameters, $\theta_2^2 = \theta_3^2$ is also where the spectrum in the gap closes for $k=\pi$ as expected [see Fig.~\ref{SIfig:fig1}].


\section{Analytics: Chiral currents in two-dimensional rectangular lattice}

In this section, we analyze a simple model introduced in Ref.~\cite{tang2021topology} for biopolymer formation.
In this model, biopolymers can contain different numbers of two monomers $X$ and $Y$ and be in one of four different internal states $z=A,B,C,D$. As shown in Fig.~\ref{Fig:fig2}, these internal states tag a monomer and prime the biopolymer for the addition or removal of a subunit.
  
The state of the system is determined by the ordered pair $(x,y)$ that represents the number of each monomer incorporated into the biopolymer and an internal variable $w=A,B,C,D$ that encodes the internal state. We assume that the biopolymer can have at most $N_x$ of the $X$ monomers and $N_y$ of the $Y$ monomers so that  $0\le x\le N_x$ and $0\le y \le N_y$. This basic dynamics can be visualized as a lattice where edges correspond to reactions and nodes to different states of the biopolymer. 

Let us denote by $c_i$ the number of polymers in state $i$ (where the index $i=(x,y,z)$ refers to an ordered triplet specifying the state of the biopolymer). We know that the dynamics of this system can also be specified by a stoichiometric matrix $S_{i \alpha}$ where $i$ runs over biopolymer states (i.e., nodes in the lattice) and $\alpha$ runs over possible reactions (i.e., edges in the lattice). In writing $S_{i \alpha}$, we must make a gauge choice for which direction in each reaction we call forward and which reaction we call backward. We choose the convention shown in Fig.~\ref{Fig:fig2}. With this choice, we have
\be
S_{i \alpha}= \begin{cases}
1\text{ if reaction } \alpha \text{ ``produces'' a biopolymer in state } i\\
-1 \text{ if reaction }\alpha \text{ ``consumes'' a biopolymer in state } i \\
0 \text{ otherwise} 
\end{cases}
\ee
and the full dynamics are given by
\be
{d c_i \over dt} = \sum_{\alpha} S_{i \alpha}J_\alpha,
\ee
where $J_\alpha$ is the net flux across reaction $\alpha$.

It is clear from Fig.~\ref{Fig:fig2} that $S_{i\alpha}$ is periodic in both the $x$- and $y$-directions. The elementary lattice cell consists of four chemical species corresponding to the four internal states $z=A,B,C,D$ at a fixed $(x,y)$ and eight elementary reactions corresponding to the four internal reactions (modifying the state $z$) and four external reactions (adding or removing a monomer of $X$ or $Y$). Translations in $X$($Y$) correspond to adding/removing a monomer of type $X$($Y$). For these reasons it is natural to consider the Fourier transform of this stoichiometric matrix,
 \be
\hat{S}(k)=\mqty[-1 & 1  & 0  & 0  & e^{ik_x \over 2}    & 0                    & -e^{ik_y \over 2}   & 0 \\
0  & -1 & 1  & 0  & 0                   &  -e^{ik_x \over 2}   & e^{-{ik_y \over 2}} & 0 \\
0  & 0  & -1 & 1  & 0                   & e^{-{ik_x \over 2}} & 0                   & -e^{-{ik_y \over 2}}\\
1  & 0  & 0  & -1 & -e^{-{ik_x \over 2}}& 0                    & 0                   &  e^{ik_y \over 2}],
 \ee
where $e^{ik_x}$ encodes shifts in the $x$-direction and $e^{ik_y}$ encode shifts in the $y$-direction.

As before, instead of writing our dynamics in terms of the \emph{net flux} $J_\alpha$, it is useful to write the flux in terms of the backwards flux $J_\alpha^-$. 
Substituting Eq.~\eqref{SIJchem} into Eq.~\eqref{SI:steady-state-eq} yields the modified steady-state
equation
\be
\sum_{\alpha} S_{i \alpha}(e^{-{\Delta \mu_\alpha \over k_B T}} -1)J_\alpha^- = 0. 
\ee
Defining the parameter
\be
\theta_\alpha\equiv e^{-{\Delta \mu_\alpha \over k_B T}} -1,
\ee
yields the chemical potential-dependent stoichiometric matrix 
\be
S_{i \alpha}^- \equiv S_{i \alpha} \theta_\alpha = S_{i \alpha}(e^{-{\Delta \mu_\alpha \over k_B T}} -1).
\label{SI:defSminus}
\ee
In Fourier space, this becomes
\be
\hat{S}^-(k)=\mqty[-\theta_1 & \theta_2  & 0  & 0  & \theta_5 e^{ik_x \over 2}    & 0                    & -\theta_7e^{ik_y \over 2}   & 0 \\
0  & -\theta_2 & \theta_3  & 0  & 0                   &  -\theta_6 e^{ik_x \over 2}   & \theta_7 e^{-{ik_y \over 2}} & 0 \\
0  & 0  & -\theta_3 & \theta_4  & 0                   & \theta_6 e^{-{ik_x \over 2}} & 0                   & -\theta_8 e^{-{ik_y \over 2}}\\
\theta_1  & 0  & 0  & -\theta_4 & -\theta_5 e^{-{ik_x \over 2}}& 0                    & 0                   &  \theta_8 e^{ik_y \over 2} ].
\ee

In order to identify a topological invariant, we examine the spectrum of 
\be
\hat{H}(k)= \hat{S}^-(k) \hat{S}^-(k)^\dagger= \mqty[\theta_1^2+\theta_2^2+\theta_5^2+\theta_7^2 & -\theta_2^2-\theta_7^2e^{ik_y} & 0 &-\theta_1^2-\theta_5^2e^{ik_x}\\
 -\theta_2^2-\theta_7^2e^{-ik_y}& \theta_2^2+\theta_3^2+\theta_5^2+\theta_7^2 & -\theta_3^2-\theta_6^2e^{ik_x} &
 0 \\
 0 &-\theta_3^2-\theta_6^2e^{-ik_x}&\theta_3^2+\theta_4^2+\theta_6^2+\theta_8^2 &-\theta_4^2-\theta_8^2e^{-ik_y}\\
 -\theta_1^2-\theta_5^2e^{-ik_x}&0&-\theta_4^2-\theta_8^2e^{ik_y}& \theta_1^2+\theta_4^2+\theta_5^2+\theta_8^2].
\ee

We now focus on the case when the ``internal'' reactions are identical so that $\Delta \mu^0_\alpha + \drive_\alpha=  a$ for $\alpha=1,2,3,4$ and the ``external'' reactions are identically driven so that  $\Delta \mu^0_\alpha + \drive_\alpha=  b$ for $\alpha=5,6,7,8$. In this case, translational and 4-fold rotational symmetry dictate that the concentrations for a periodic lattice  must all be uniform so that the thermodynamic drives have the same symmetry as the driving forces with two corresponding values, $\theta_a$ and $\theta_b$.
This gives us
\be
\hat{H}(k)= \hat{S}^-(k) \hat{S}^-(k)^\dagger=\mqty[2\theta_a^2 +2\theta_b^2& -\theta_a^2-\theta_b^2e^{ik_y} & 0 &-\theta_a^2-\theta_b^2e^{ik_x}\\
 -\theta_a^2-\theta_b^2e^{-ik_y}& 2\theta_a^2+2\theta_b^2 & -\theta_a^2-\theta_b^2e^{ik_x} &
 0 \\
 0 &-\theta_a^2-\theta_b^2e^{-ik_x}&2\theta_a^2+2\theta_b^2 &-\theta_a^2-\theta_b^2e^{-ik_y}\\
 -\theta_a^2-\theta_b^2e^{-ik_x}&0&-\theta_a^2-\theta_b^2e^{ik_y}&2 \theta_a^2+2\theta_b^2],
\ee
where $\theta_a=e^{-{a \over k_B T}}-1$ and $\theta_b=e^{-{b \over k_B T}}-1$.
As before, it is helpful to separate out the $k$-dependent and independent parts,
\be
\hat{H}(k)=H_0- \hat{H}_1(k),
\ee
with
\be
H_0= \mqty[2\theta_a^2 +2\theta_b^2&0 & 0 &0\\
0& 2\theta_a^2+2\theta_b^2 &0 &0 \\
 0 &0 &2\theta_a^2+2\theta_b^2 &0\\
 0&0&0&2 \theta_a^2+2\theta_b^2],
\ee
and
\be
\hat{H}_1(k)= \mqty[ 0& \theta_a^2+\theta_b^2e^{ik_y} & 0 &\theta_a^2+\theta_b^2e^{ik_x}\\
 \theta_a^2+\theta_b^2e^{-ik_y}& 0 & \theta_a^2+\theta_b^2e^{ik_x} &
 0 \\
 0 &\theta_a^2+\theta_b^2e^{-ik_x}&0 &\theta_a^2+\theta_b^2e^{-ik_y}\\
 \theta_a^2+\theta_b^2e^{-ik_x}&0& \theta_a^2+\theta_b^2e^{ik_y}&0].
\ee
We can then rewrite this as
\be
\hat{H}_1(k)=I\otimes [\vec{a}(k_y)\cdot \vec{\sigma}]+\sigma_x \otimes [\vec{a}(k_x)\cdot \vec{\sigma}],
\ee
where $\vec{a}=(\theta_a^2+\theta_b^2 \cos{x}, \theta_b^2 \sin{x},0)$ and $\vec{\sigma}=(\sigma_x, \sigma_y, \sigma_z)$ is the vector of the Pauli matrices. This suggests that in analogy with the triangular lattice, there should be a topological phase transition when $\theta_a^2=\theta_b^2$, the gap closes, and $\vec{a}(k_x)$ and $\vec{a}(k_y)$ go from enclosing the origin to not enclosing the origin.

This basic intuition can be verified by directly diagonalizing $\hat{H}_1(k)$ and looking at the corresponding eigenvalues and eigenvectors. To do so, it is useful to define 
\be
A(k)= \theta_a^2+\theta_b^2e^{ik}= |A(k)|e^{i \phi(k)},
\ee
where $|A(k)|=\sqrt{\theta_a^4+\theta_b^4+2\theta_a^2\theta_b^2 \cos{k}}$ is just the magnitude and $\phi_k$ is the corresponding phase. In terms of these, we can write the four eigenvalues as
\bea
\lambda= \left\{\pm \mathrm{abs}\left(|A(k_y)|^2-|A(k_x)|^2\right), \pm \left(|A(k_y)|^2+|A(k_x)|^2\right)\right\}.
\eea
We see that the spectrum becomes degenerate when $|A(k_x)|^2=0$  or  $|A(k_y)|^2=0$  which occur when
$\theta_a^2=\theta_b^2$ and $k_x=\pi$ or $k_y=\pi$, respectively. We can also look at the corresponding eigenvectors. To do so, we define a variable $d=\mathrm{sgn}(\left(|A(k_y)|^2-|A(k_x)|^2\right)$ that takes on values $\pm 1$. In terms of these quantities, one can show that the four eigenvectors are
\bea
|e_1\> &=& \qty[de^{i\phi(k_x)}, -e^{i[\phi(k_x)-\phi(k_y)]},-de^{-i\phi(k_y)},1] \nonumber \\
|e_2\> &=& \qty[-de^{i\phi(k_x)}, -e^{i[\phi(k_x)-\phi(k_y)]},de^{-i\phi(k_y)},1] \nonumber \\
|e_3\> &=& \qty[-e^{i\phi(k_x)}, e^{i[\phi(k_x)-\phi(k_y)]},-e^{-i\phi(k_y)},1] \nonumber \\
|e_4\> &=& \qty[e^{i\phi(k_x)}, e^{i[\phi(k_x)-\phi(k_y)]},e^{-i\phi(k_y)},1].
\eea
The corresponding gauge potential (i.e., Berry phase) for the first eigenstate is just
\be
(A_{kx},A_{ky})=(\<e_1|\partial_{k_x} e_1\>,\<e_1|\partial_{k_y} e_1\>)=2i\qty({\partial \phi(k_x) \over \partial k_x}, {\partial \phi(k_y) \over \partial k_y}).
\ee
We can then look at the corresponding charge over a Wilson loop (contour integral in $k_x$ and $k_y$ space) to define the $\mathbb{Z}_2$ topological invariant. We see that when $\theta_a^2 \ge \theta_b^2$, this phase is trivial in both cases (the origin is not enclosed), whereas for $\theta_a^2 \le \theta_b^2$ it is nontrivial and encloses origin. This confirms that there is a topological phase transition when $\theta_a^2=\theta_b^2$.

The exact nature of the phase where topological edge currents are supported can be gleaned by numerical simulations. We find that edge currents exist in one of these phases, but not the other [see Fig.~\ref{Fig:fig1}].


\section{Numerical Simulations}

In this section, we provide details on numerical simulations. All code used to perform simulations and generate figures can be found in the corresponding Jupyter Notebook on our GitHub repository at  \url{https://github.com/Emergent-Behaviors-in-Biology/NESS-Topology}.

\subsection{One-dimensional triangular lattice}
In order to simulate the triangular lattice, we directly simulate the ODE system defined by Eqs.~\eqref{Eq:dynamics1} and \eqref{Eq:massaction} using ODEint from SciPY package ~\cite{virtanen2020scipy}.
We specify the six kinetic constants corresponding to the forward and backward rates for the three types of reactions in the lattice cells as $k_1^\pm$, $k_2^\pm$, and $k_3^\pm$. For all three reactions,
we choose backward reactions so that $k_1^-=k_2^-=k_3^-=1$. Furthermore, we choose forward rates $k_j^+$ to be thermodynamically consistent using Eq.~\eqref{SIEq:k_+k_-} so that
\be
k_A^+=k_A^- e^{-{\Delta \mu_A^0 +\drive_A \over k_B T}},
\ee
for $A=1,2,3$. As discussed in main text, we choose the first reaction to be at equilibrium
so that
\be
\Delta \mu_1^0 +\drive_1=0
\ee
and vary $\Delta \mu_A^0 +\drive_A$ for reactions 2 and 3 as indicated in the main text. We choose initial conditions by sampling $c_i$ uniformly in the interval $[0, 5]$ (though any reasonable initial conditions for which the integrator converges give identical results).
We compute fluxes, chemical potentials, and thermodynamic drives directly from the steady-state concentrations. See Jupyter Notebook for detailed code.

\subsection{Two-dimensional rectangular lattice}
In order to simulate the two-dimensional rectangular lattice, we make use of the Gillespie algorithm for simulating chemical reaction networks ~\cite{gillespie1976general}. Direct numerical simulation of the corresponding ODEs proved to be numerically unstable due to positivity constraints. In addition to the stoichiometric matrix, simulations require
us to specify the 16 kinetic constants corresponding to the 8 types of reactions that can occur. Following Tang et. al. ~\cite{tang2021topology}, we divide this 8 reactions into two classes: 4 ``external reactions'' that change the  number of monomers [reactions 5-8 in Fig.~\ref{Fig:fig2}] and 4 ``internal reactions'' [reactions 1-4 in Fig.~\ref{Fig:fig2}] that change the internal states of the biopolymer.

We choose $k_j^-=0.01$ ($j=1,\ldots,8$) for all 8 reactions and use Eq.~\eqref{SIEq:k_+k_-} to 
set
\be
k_j^+=e^{-{\Delta \mu_a^0 +\drive_a \over k_B T}}k_j^-
\ee
for reactions $j=1,2,3,4$ and
and 
\be
k_j^+=e^{-{\Delta \mu_b^0 +\drive_b \over k_B T}} k_j^-
\ee
for reactions $j=5,6,7,8$. The two parameters $\Delta \mu_a^0 +\drive_a$ and $\Delta \mu_b^0 +\drive_b$ are chosen such that  $k_a^-= k_b^-=0.01$, $k_b^+=0.03$ and $k_a^+$ is varied.

To calculate the average abundance of a chemical species $c_i$ we average the stochastic trajectory corresponding to the species. The average flux is then calculated directly from the simulated average abundances using the formula for the currents and mass action kinetics.
See Jupyter Notebook for detailed code.

\newpage
\section{Supplementary Figures}


\end{document}